\documentclass[%
 reprint,
 amsmath,amssymb,
 aps,
 nofootinbib
]{revtex4-2}

\usepackage{graphicx}
\usepackage{dcolumn}
\usepackage{bm}
\usepackage{tabularx}
\usepackage{multirow}
\usepackage{xcolor}
\usepackage{hyperref}
\newcommand{\SC}{\ensuremath{{\rm SC}}}
\newcommand{\NSC}{\ensuremath{{\rm NSC}}}
\newcommand{\AC}{\ensuremath{{\rm AC}}}
\newcommand{\NAC}{\ensuremath{{\rm NAC}}}
\newcommand{\llangle}{\ensuremath{\langle\langle}}
\newcommand{\rrangle}{\ensuremath{\rangle\rangle}}

\newcommand{\rlangle}{\ensuremath{\rangle\langle}}

\begin{document}

\preprint{APS/123-QED}

\title{Flow harmonic correlations via multi-particle symmetric and asymmetric cumulants in Au+Au collisions at \(\sqrt{s_{NN}}\) = 200 GeV}

\author{Kaiser Shafi}
\email{kaisers@iiserbpr.ac.in}
\author{Prabhupada Dixit}
\email{prabhupadad@iiserbpr.ac.in}
\author{Sandeep Chatterjee}
\email{sandeep@iiserbpr.ac.in}
\author{Md. Nasim}
\email{nasim@iiserbpr.ac.in}

\affiliation{Department of Physical Sciences, Indian Institute of Science Education and Research Berhampur,\\ Laudigam--760003, Dist.--Ganjam, Odisha, India}
\date{\today}

\begin{abstract}
We study multi-particle azimuthal correlations in Au+Au collisions at $\sqrt{s_{NN}}$ = 200 GeV. We use initial conditions obtained from a Monte Carlo Glauber model and evolve them within a viscous relativistic hydrodynamics framework that eventually gives way to a transport model in the late hadronic stage of the evolution. We compute the multi-particle symmetric and asymmetric cumulants and present the results for their sensitivity to the shear and bulk viscosities during the hydrodynamic evolution. We also check their sensitivity to resonance decay and hadronic interactions. We demonstrate that while some of these observables are more sensitive to transport properties than traditional flow observables, others are less sensitive, making them suitable for studying different stages of the evolution. 
\end{abstract}

\maketitle

\section{Introduction}
In ultra-relativistic heavy-ion collisions, a thermalized, strongly interacting, deconfined medium, known as Quark-Gluon Plasma (QGP), is expected to form~\cite{Collins:1974ky,Shuryak:2004cy, Busza:2018rrf, STAR:2000ekf,STAR:2007mum,PHENIX:2004vcz}. The information regarding certain properties of QGP is carried by anisotropies observed in the momentum distributions of the produced particles~\cite{Heinz:2013th,Braun-Munzinger:2015hba,Busza:2018rrf}. After hadronization, particles are emitted anisotropically in the transverse plane to the beam direction, as detected by a detector. Traditionally, Fourier series in flow amplitudes $v_n$ and symmetry planes $\psi_n$ are used to quantify this azimuthal distribution in the momentum space~\cite{Voloshin:1994mz},
\begin{equation}
     \frac{dN}{d\phi}= \frac{1}{2\pi} \left[ 1 + 2 \sum_{n=1}^{\infty} v_n \cos\left[n\left(\phi-\psi_n\right)\right] \right]
\end{equation}
\par
Collective flow is crucial for studying the properties of the QGP medium formed in these collisions~\cite{Ollitrault:1992bk, Heinz:2013th,Bass:1998vz}. The impact parameter driven spatial geometry of the fireball, event-by-event nucleon position as well as sub-nucleonic fluctuations within the overlap region of the two colliding nuclei lead to the development of various orders of anisotropic flow harmonics. Due to the almond shape of the overlap region, the primary source of the second-order flow harmonic, known as elliptic flow ($v_{2}$), is the initial spatial anisotropy~\cite{Voloshin:2008dg, Voloshin:1999gs, Ivanov:2014zqa, STAR:2004jwm, STAR:2001ksn,ALICE:2010suc,Sorge:1998mk}. The third-order flow harmonic, known as triangular flow ($v_{3}$), arises from event-by-event fluctuations in the positions of the colliding nucleons as well as their sub-nucleonic fluctuations~\cite{STAR:2013qio,Solanki:2012ne,ALICE:2016cti,Heinz:2013bua, Alver:2010gr}. These fluctuations also give rise to higher-order flow harmonics. Previous studies have demonstrated that these flow harmonics are sensitive to the equation of state (EoS) and the transport properties of the fireball created in the collision, such as shear and bulk viscosity~\cite{Parkkila:2021tqq,Retinskaya:2014zea,Shen:2012vn,Teaney:2003kp,Shen:2011zc}.
\par
Anisotropic flow analysis involves measuring $v_n$, $\psi_n$, and their event-by-event correlations and fluctuations. Due to the random fluctuations of the impact parameter vector, however, the estimation of $v_n$ and $\psi_n$ involves the complications associated with reaction plane dispersion in conventional flow analyses. As a result, they are indirectly estimated using correlation techniques, in which there is no need for an event-by-event determination of the reaction plane~\cite{Wang:1991qh,Jiang:1992bw}. Therefore, there is no need to correct for dispersion in an estimated reaction plane. The foundation of this alternative approach is based on the following outcome,
\begin{equation}
\label{foundation_eq}
    \langle e^{i\left(n_1\phi_1+n_2\phi_2+...+n_k\phi_k\right)} \rangle = v_1v_2...v_k e^{i\left(n_1\psi_1+n_2\psi_2+...+n_k\psi_k\right)}
\end{equation}
which analytically relates multi-particle azimuthal correlators and flow degrees of freedom~\cite{Bhalerao:2011yg}. The average is calculated over all unique sets of $k$ different particles in a single event. This expression can be used to determine the properties of flow amplitudes $v_n$ and symmetry planes $\psi_n$ on an event-by-event basis. Apart from collective flow, other sources of correlations called non-flow are also present that typically involve only a subset of particles.
\par
Multivariate cumulants were introduced in anisotropic flow analyses in the early 2000s in Refs.~\cite{PhysRevC.63.054906, PhysRevC.64.054901}, which tackled long-standing issues in the field and transformed the approach to anisotropic flow analysis in high-energy physics. For example, the four-variate cumulant that is used to estimate the flow amplitude, $v_n$, from four-particle correlation is defined as~\cite{PhysRevC.64.054901}
\begin{eqnarray}
\label{Eq-1}
    c_{n}\{4\} = \langle\langle e^{in(\phi_{1} + \phi_{2} - \phi_{3} - \phi_{4})} \rangle\rangle - 2\langle \langle e^{in(\phi_{1} - \phi_{3})} \rangle \rangle^{2},
\end{eqnarray}
The double angular brackets indicate that the averaging procedure is performed in two steps, averaging over all distinct particle multiplets in an event and then averaging these single-event averages with appropriately chosen event weights. By generalizing this concept for non-identical harmonics, new observables, which strictly satisfy all defining mathematical properties of cumulants, were constructed to quantify the correlations among different flow harmonics~\cite{Bilandzic:2013kga}. Ref.~\cite{PhysRevC.105.024912} further generalized these observables to probe the genuine correlation between different moments of flow harmonics.

\subsection{Symmetric Cumulants of Flow Amplitudes}
Reference~\cite{Bilandzic:2013kga} proposed a general algorithm to measure multi-particle correlation where the harmonics inside the correlator bracket of Eq. (\ref{Eq-1}) can be different. This introduces a new set of observables, known as symmetric cumulants $(SCs)$, which can be used to measure the correlation between event-by-event fluctuation of flow harmonics $v_{m}$ and $v_{n}$. This new approach allows for the separation of non-flow and flow contributions and provides initial insights into how the combinatorial background contributes to flow measurements using correlation techniques. These four-particle symmetric cumulants, $\SC(m,n)$, are defined as~\cite{Bilandzic:2013kga}
\begin{eqnarray}
\SC(m,n) &\equiv& \langle v_m^2 v_n^2 \rangle _c \nonumber\\
        &=& \left<v_{m}^2v_{n}^2\right>-\left<v_{m}^2\right>\left<v_{n}^2\right>\nonumber\\
        &=& \left<\left<\cos(m\varphi_1\!+\!n\varphi_2\!-\!m\varphi_3-\!n\varphi_4)\right>\right>\nonumber\\
        && - \left<\left<\cos[m(\varphi_1\!-\!\varphi_2)]\right>\right>\left<\left<\cos[n(\varphi_1\!-\!\varphi_2)]\right>\right>\nonumber\\
\label{scmn_1}
\end{eqnarray}
The subscript $c$ indicates the cumulant. Positive (Negative) values of $\SC(m,n)$ suggest the correlation (anti-correlation) between  $v_{m}^{2}$ and $v_{n}^{2}$, which means that if $v_{m}^{2}$ is larger than $\langle v_{m}^{2}\rangle$ in an event then the probability of $v_{n}^{2}$  being larger than $\langle v_{n}^{2}\rangle$ in that same event is enhanced (suppressed).  The $\SC(m,n)$ observables focus on the correlations among different orders of flow harmonics and enable the quantitative comparison between experimental data and model calculations. To eliminate the effect of the magnitudes of $v_m$ and $v_n$ on the value of the symmetric cumulant, we divide $\SC(m,n)$ by their average values, $\left<v_m^2\right>$ and $\left<v_n^2\right>$, and define the normalized symmetric cumulant. This enables us to compare data and model calculations in a quantitative way and compare the fluctuations of the initial and final states. The normalized symmetric cumulant, denoted by $\NSC(m,n)$, is achieved following the standard method from Ref.~\cite{Taghavi:2020gcy}:
\begin{equation}
\NSC(m,n)= \frac{\SC(m,n)}{\langle
  v_{m}^{2}\rangle  \langle v_{n}^{2} \rangle}. 
\label{scmn_norm1}
\end{equation}

These correlations have been measured by both STAR experiment at RHIC~\cite{STAR:2018fpo} and ALICE experiment at LHC~\cite{ALICE:2016kpq} with the observation of a positive correlation between $v_{2}$ and $v_{4}$ while a negative correlation between $v_{2}$ and $v_{3}$. The sensitivity of these observables to the shear viscosity to entropy density ($\eta/s$) has been studied in transport models like AMPT~\cite{Nasim:2016rfv}.

\subsection{Asymmetric Cumulants of Flow Amplitudes}
Recent studies indicate that insightful information can be obtained about the properties of Quark-Gluon Plasma by using higher-order observables~\cite{Parkkila:2021yha}. These observables can probe the genuine correlations between the different moments of different flow harmonics. They are robust against non-flow correlations, which can be verified using the HIJING Monte Carlo generator~\cite{Mordasini:2019hut}. The asymmetric cumulant $\AC_{2,1}(m,n)$ is defined as~\cite{PhysRevC.105.024912}
\begin{equation}
    \begin{split}
    \label{eq_ac21}
        \AC_{2,1}(m,n) & \quad \equiv \langle (v_m^2)^2 v_n^2 \rangle _c
         \quad \equiv \langle v_m^4 v_n^2 \rangle _c \\
        & \quad = \langle v_m^4 v_n^2 \rangle - \langle v_m^4 \rlangle v_n^2 \rangle - 2 \langle v_m^2 v_n^2 \rlangle v_m^2 \rangle \\
        & \quad + 2 \langle v_m^2 \rangle^2 \langle v_n^2 \rangle
    \end{split}
\end{equation}
In Eq. (\ref{eq_ac21}), the subscript (2,1) on the left-hand side indicates the exponents of $v_m^2$ and $v_n^2$ on the right-hand side. The combinations of azimuthal correlators used to estimate the $ACs$ are~\cite{PhysRevC.105.024912}:
\begin{equation}
    \begin{split}
        \AC_{2,1}(m,n) & = \llangle {\rm e}^{i(m\varphi_1 + m\varphi_2 + n\varphi_3 - m\varphi_4 - m\varphi_5 - n\varphi_6)} \rrangle\\
        & \quad - \llangle {\rm e}^{i(m\varphi_1 + m\varphi_2 - m\varphi_3 -m\varphi_4)} \rrangle \llangle {\rm e}^{i(n\varphi_1 - n\varphi_2)} \rrangle \\
        & \quad - 2  \llangle {\rm e}^{i(m\varphi_1 + n\varphi_2 - m\varphi_3 - n\varphi_4)} \rrangle \llangle {\rm e}^{i(m\varphi_1 - m\varphi_2)} \rrangle \\
        & \quad + 2  \llangle {\rm e}^{i(m\varphi_1 - m\varphi_2)} \rrangle^2 \llangle {\rm e}^{i(n\varphi_1 - n\varphi_2)} \rrangle
    \end{split}
\end{equation}
These expressions are genuine multivariate cumulants. $\AC_{1,1}(m,n)$ corresponds to SC$(m,n)$, illustrating the generalization aspect of the $ACs$. We can also normalize the $ACs$. This procedure has two benefits. Normalizing the results allows proper comparisons and determination of the initial state effects and the changes brought by the hydrodynamic evolution since the predictions for $ACs$ in the initial and final state do not have the same scale. Flow amplitudes depend on the transverse momentum, $p_T$, which leads to a similar dependence in any linear combinations of them, such as the $SCs$ and the $ACs$. The normalization eliminates this dependence and enables comparisons between models and data with different $p_T$ ranges~\cite{Mordasini:2019hut}. Again, the normalization of the $ACs$ is done following the standard method from Ref.~\cite{Taghavi:2020gcy},
\begin{eqnarray}
\label{NACDefinition}
\NAC_{2,1}(m,n) & = & \frac{{\AC}_{2,1}(m,n)}{\langle v_m^2 \rangle^2 \langle v_n^2 \rangle}.
\end{eqnarray}

\par
In this paper, we have computed $\SC(m,n)$ and $\AC_{2,1}(m,n)$ in Au+Au collisions at $\sqrt{s_{NN}} = 200$ GeV using a hybrid model. $\SC(m,n)$ have been computed in previous studies from hydrodynamic and transport models~\cite{Zhu:2016puf, Schenke:2019ruo, Mordasini:2019hut, Taghavi:2020gcy, Hirvonen:2022xfv, Magdy:2022ize, Magdy:2024ooh, Gardim:2016nrr}, while $\AC_{2,1}(m,n)$ have been computed in transport models~\cite{Magdy:2022ize, Magdy:2024ooh}. The ALICE collaboration has performed a systematic study of $\SC(m,n)$ and $\NSC(m,n)$~\cite{ALICE:2017kwu}. The ALICE collaboration has also measured $\AC_{2,1}(m,n)$ for Pb+Pb collisions at $\sqrt{s_{NN}}=5.02 \ TeV$~\cite{ALICE:2023lwx}. In addition to symmetric cumulants, various other correlators have also been studied~\cite{ATLAS:2018ngv,Ortiz:2019osu,CMS:2019lin,ALICE:2021adw}. We show the sensitivity of these measurements to the transport properties, such as the shear viscosity to entropy density ratio, $\eta/s$, and the bulk viscosity to entropy density ratio, $\zeta/s$, of the medium. 

\section{Framework}
\label{sec-1}
We use a framework with multiple components to simulate various stages of heavy ion collisions. The hydrodynamic evolution has been initialized using a Glauber-based model. To generate the incoming nuclei, we use a Woods-Saxon distribution to sample nucleons.
\begin{equation}
    \rho = \frac{\rho_0}{1 + exp[(r-R)/a]}.
\end{equation}
Here, $r=\sqrt{x^2+y^2+z^2}$, $a$ characterizes the diffuseness of the nuclear surface, $R$ is the radius parameter, and $\rho_0$  is the saturation density determined by $\int \rho \ dr^3=A$ (mass number of the nucleus). We have taken $R = 6.38$ fm, and $a = 0.535$ fm~\cite{Shou:2014eya}.
\par
The evolution of the energy-momentum tensor starts at $\tau_0=0.6$ fm. The hydrodynamic equations are solved using the MUSIC simulation~\cite{Schenke:2010nt, Schenke:2010rr, Paquet:2015lta, Huovinen:2012is}, which utilizes a Kurganov-Tadmor algorithm. A constant effective $\eta/s$ = 0.08 was fixed by reproducing the measured anisotropic flow coefficients of charged hadrons. We use a temperature-dependent specific bulk viscosity parametrized in the following way~\cite{Denicol:2009pe}:
\begin{equation}\label{eqn:param_zetabys}
\frac{\zeta }{s}=\left\{ 
\begin{array}{ll}
\lambda _{1}e^{-(x-1)/\sigma _{1}}+\lambda _{2}e^{-(x-1)/\sigma _{2}}+0.001
& (T>1.05T_{C}) \\ 
\lambda _{3}e^{(x-1)/\sigma _{3}}+\lambda _{4}e^{(x-1)/\sigma _{4}}+0.03 & 
(T<0.995T_{C})
\end{array}
\right.\nonumber
\end{equation}
where $x=T/T_{C}$. The fitted parameters are $\lambda _{1}=\lambda _{3}=0.9$
, $\lambda _{2}=0.25$, $\lambda _{4}=0.22$, $\sigma _{1}=0.025$, $\sigma_{2}=0.13,$
$\sigma _{3}=0.0025$, $\sigma _{4}=0.022,$ $A_{1}=-13.77$,
$A_{2}=27.55$ and $A_{3}=13.45$.
\par
We use a QCD equation of state (EoS), NEoS-B~\cite{Monnai:2019hkn}, based on continuum extrapolated lattice calculations at zero net baryon chemical potential published by the HotQCD Collaboration~\cite{HotQCD:2014kol,HotQCD:2012fhj,Ding:2015fca}. It is smoothly matched to a hadron resonance gas EoS in the temperature region between 110 and 130 MeV~\cite{Moreland:2015dvc}.
\par
A hyper-surface is generated from the hydrodynamic space-time evolution of the fluid. To describe the dilute hadronic phase, we utilize the iSS code~\cite{Shen:2014vra,iSS:chunshen} to sample the primordial hadrons from the hypersurface with a constant energy density, $e_{sw} = 0.26 \ GeV/fm^3$, equivalent to a local temperature of approximately 151 MeV. Then, we use the UrQMD code~\cite{Bleicher:1999xi,Bass:1998ca} to simulate scatterings and decays of hadrons during the late stage of heavy ion collisions. To increase statistics, each hydrodynamic switching hyper-surface is sampled multiple times. The number of oversampling events for every hydrodynamic event is determined to ensure sufficient statistics for every centrality class. We generated ensembles of 2000 hydrodynamic events for each centrality class, each with a suitable number of repeated samplings. Table~\ref{table:nEvents} lists the number of events generated for each of the centrality classes with the three sets of parameters for the hydrodynamic stage shown in Table~\ref{table:parsets}. Table~\ref{table:stages} outlines the particles used in the analyses at the end of different stages of the evolution. Analysis at the end of: \\
$\bullet$ stage I includes hydrodynamic evolution and particlization,\\
$\bullet$ stage II includes hydrodynamic evolution, particlization, and resonance decay, and\\
$\bullet$ stage III includes hydrodynamic evolution, particlization, resonance decay, and hadronic interactions within the UrQMD model approach.

\begin{table}[t!]
\centering
\begin{tabular}{ c|c|c|c|c|c| }
\hline
\multicolumn{1}{|c|}{Centrality class} & 0-10\% & 10-20\% & 20-30\% & 30-40\% & 40-50\% \\
 \hline
\multicolumn{1}{|c|}{\# Events \scriptsize{(x$10^6$)}} & 4 & 6 & 8 & 3 & 4 \\
 \hline
\end{tabular}
\caption{Ensemble size (in millions) of the centrality classes at the end of Stage III.}
\label{table:nEvents}
\end{table}

\begin{table}[t!]
\centering
\begin{tabular}{ |c|c|c| }
\cline{1-3}
  Parameter(Par.) Sets & $\eta/s$ & $\zeta/s$ \\
 \hline
\multicolumn{1}{|c|}{I} & 0 & 0 \\
 \hline
 \multicolumn{1}{|c|}{II} & 0.08 & 0\\ 
 \hline
\multicolumn{1}{|c|}{III} & 0.08 & $\zeta/s(T)$\\
 \hline
\end{tabular}
\caption{The transport coefficients' values for the parameter sets used in the ensembles generated for this study.}
\label{table:parsets}
\end{table}

\begin{table}[t!]
\centering
\begin{tabular}{ |c|c| }
\cline{1-2}
 Stages & Particle list \\
 \hline
\multicolumn{1}{|c|}{I} & Hydrodynamic evolution + particlization (Primordial) \\
 \hline
 \multicolumn{1}{|c|}{II} & Primordial + Resonance decays\\ 
 \hline
\multicolumn{1}{|c|}{III} & Primordial + UrQMD\\
 \hline
\end{tabular}
\caption{The particle lists analyzed at the end of different stages of the evolution (Hydrodynamic evolution, resonance decay, and hadronic transport).}
\label{table:stages}
\end{table}

\par
In the following results, some observables will be presented by scaling with the average number of participants, $\langle N_{part} \rangle$. Table~\ref{table:avnpart} lists the values of $\langle N_{part} \rangle$ in different centrality bins, which are determined using the Monte Carlo Glauber model. The $\SC(m,n)$ and $\AC_{2,1}(m,n)$ observables have been computed using particles with transverse momentum ($p_T$) ranges of [0.2, 2.0] GeV and [0.01, 3.0] GeV, respectively.

\begin{table}[t!]
\centering
\begin{tabularx}{0.45\textwidth}
{ | >{\centering\arraybackslash}X 
  | >{\centering\arraybackslash}X 
  | >{\centering\arraybackslash}X | }
\hline
Centrality class & $\langle N_{part}\rangle$\\
 \hline
 0-10\% & 327.7\\
 \hline
 10-20\% & 240.3\\
 \hline
 20-30\% & 172.7\\
 \hline
 30-40\% & 120.5\\
 \hline
 40-50\% & 81.5\\
 \hline
\end{tabularx}
\caption{$\langle N_{part}\rangle$ in the centrality classes from a Monte Carlo Glauber model.}
\label{table:avnpart}
\end{table}

\section{Error Estimation}
The bootstrap method is used to efficiently estimate the error on ensemble-level observables using the Monte Carlo approach, bypassing the complexities of standard error propagation. It involves randomly selecting events with replacement from an original sample to create bootstrap samples, allowing for the calculation of sampling variance.
\par
Let \(\hat{O}\) be the estimator of a statistic on which we intend to find the standard error. To estimate the standard error using the bootstrap method, we follow these steps:
\begin{itemize}
\item Given a parent sample of size \( n \), we construct \( B \) independent bootstrap samples \( X^*_1, X^*_2, \ldots, X^*_B \), each with \( n \) data points randomly drawn with replacement.
\item We evaluate the estimator for each bootstrap sample.
$$\hat{O}^*_b = \hat{O}(X^*_b), \hspace{1cm} b = 1,2,...,B.$$
\item The sampling variance of the estimator can then be calculated as follows:
$$Var(\hat{O}) =  \frac{1}{B-1}\sum_{b=1}^{B}(\hat{O}^*_b - \bar{\hat{O}}),$$
where, $\bar{\hat{O}} = \frac{1}{B}\sum_{b=1}^{B}\hat{O^*_b}$. 
\end{itemize}
The optimal value of \( B \) for accurate error estimation varies, but generally, larger values yield better estimates.

\section{Results}
We calibrated our model by using the produced particle yields,  transverse momentum spectra and elliptic flow. We then compared these results to experimental data from the PHENIX and the STAR Collaborations. The normalization factor for the system's energy density was determined by matching the model and STAR data~\cite{STAR:2008med} charged hadron multiplicity $dN_{Ch}/d\eta$ in the 0-5\% centrality bin. We also compared the $p_T$ spectra of identified particles in 0-5\% centrality Au+Au collisions to STAR results~\cite{STAR:2008med}. Additionally, we compared the single-particle anisotropic flow coefficient $v_2$ of charged particles in 20-30\% centrality Au+Au collisions with the PHENIX measurement~\cite{PHENIX:2011yyh}. The hydrodynamic evolution of the MCGM initial conditions, coupled to UrQMD, can reproduce the elliptic flow up to the mid-central collisions.

\subsection{Sensitivity to Transport Coefficients}
\begin{figure}[t!]
\includegraphics[width=0.45\textwidth]{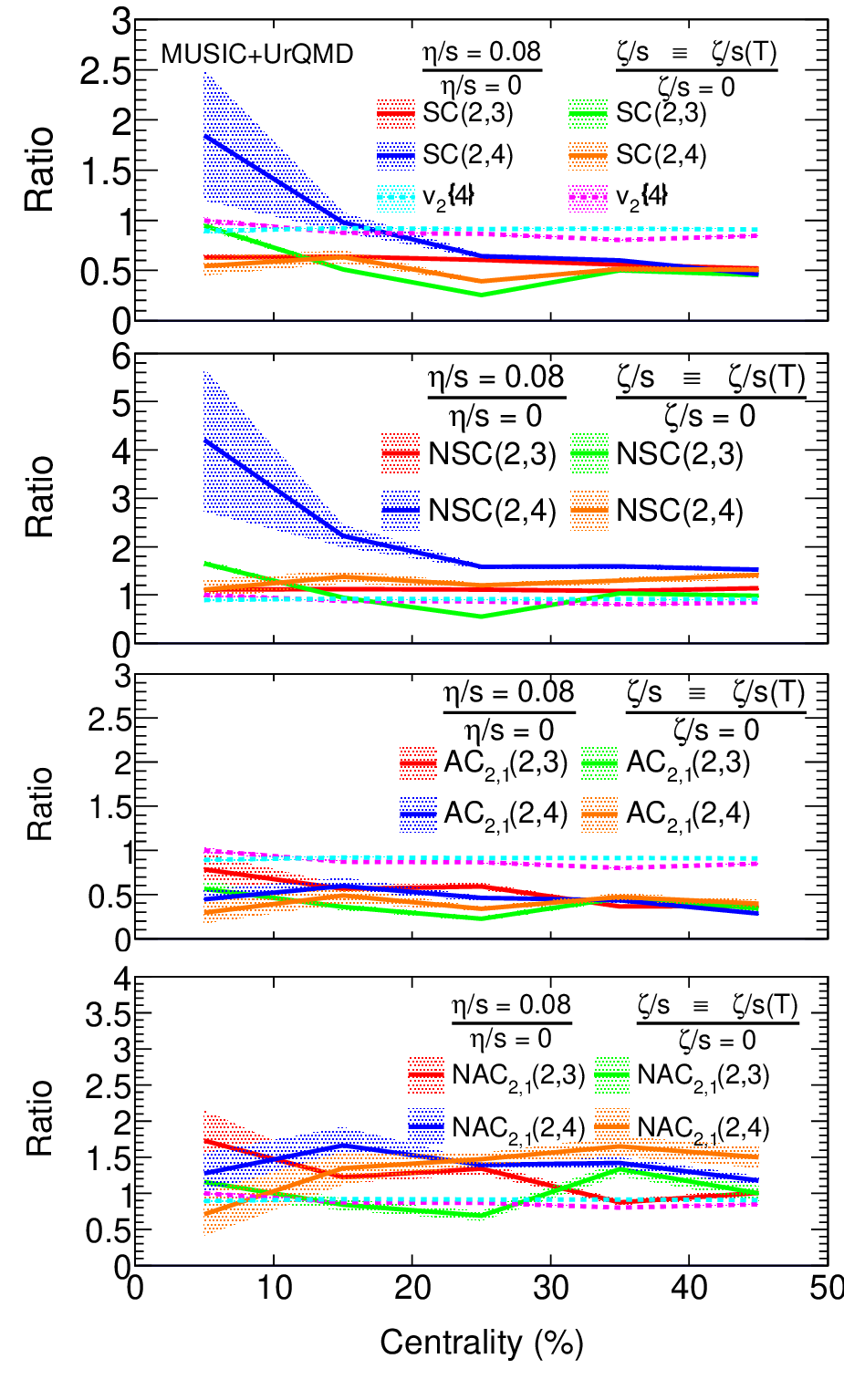}
\caption{\label{fig:ratio_plots_hydro_coeff} Ratios of $v_2\{4\}$, $\SC(2,n)$, $\NSC(2,n)$, $\AC_{2,1}(2,n)$, and $\NAC_{2,1}(2,n)$, with different hydro parameter sets, vs centrality in Au+Au collisions at $\sqrt{s_{NN}}$ = 200 GeV. The figure compares the effects of $\eta/s$ and $\zeta/s$ on $v_2\{4\}$, $\SC(2,n)$, $\NSC(2,n)$, $\AC_{2,1}(2,n)$, and $\NAC_{2,1}(2,n)$. The plotted results show ratios of $v_2\{4\}$ using cyan ($\eta/s$) and magenta ($\zeta/s$) bands and $\SC(2,n)$, $\NSC(2,n)$, $\AC_{2,1}(2,n)$, and $\NAC_{2,1}(2,n)$ using red (n=3, $\eta/s$), green (n=3, $\zeta/s$), blue (n=4, $\eta/s$) and orange (n=4, $\zeta/s$) bands. In the legends, the left column show the ratios for checking the effect of $\eta/s$ and the right column show the ratios for checking the effect of $\zeta/s$.}
\end{figure}
\subsubsection{Four-Particle Symmetric Cumulants}
In this section, we present our findings on the symmetric cumulants $\SC(m,n)$ and normalized symmetric cumulants $\NSC(m,n)$ obtained from our simulations. To evaluate their sensitivity to hydrodynamic transport coefficients, we analyzed these observables utilizing the three sets of ensembles detailed in Table~\ref{table:parsets}. 

Our results indicate that both $\SC(2,3)$ and $\SC(2,4)$ are significantly suppressed by the effects of shear and bulk viscosities. In Fig.~\ref{fig:ratio_plots_hydro_coeff}, we illustrate the influence of shear viscosity on $v_2\{4\}$ and $\SC(2,3)$. The maximum observed variation for $v_2\{4\}$ is approximately 10\% in mid-central collisions, while $\SC(2,3)$ displays a much larger change of around 40-50\% across the entire centrality range. Although the comparison with $v_{3}\{4\}$ has not been shown, it yields similar results to the comparison with $v_{2}\{4\}$, with sensitivity increasing for higher \(n\) values in $v_{n}$~\cite{Schenke:2011bn}. This trend is similarly observed with respect to the bulk viscosity of the medium, underscoring the heightened sensitivity of these observables to the transport coefficients. 

Furthermore, the impact of the transport coefficients on $\SC(2,4)$ is more pronounced than that on $\SC(2,3)$. This substantial sensitivity in symmetric cumulants arises primarily from the characteristics of anisotropic flow, as these cumulants incorporate higher powers of flow coefficients.

As discussed in the introduction, normalization serves to mitigate the influence of the magnitudes of \(v_m\) and \(v_n\) on the cumulants. Consequently, it is also expected to reduce the incidental sensitivity of \(\SC(m,n)\) to shear and bulk viscosities. Therefore, we also examined the sensitivities of the normalized cumulants. From Fig.~\ref{fig:ratio_plots_hydro_coeff}, it is apparent that $\NSC(2,3)$ exhibits minimal sensitivity to shear viscosity. However, $\NSC(2,4)$ shows greater sensitivity to both shear and bulk viscosities compared to $v_2\{4\}$.

\subsubsection{Six-Particle Asymmetric Cumulants}
Next, we present the results for the six-particle asymmetric cumulants \( \AC_{2,1}(m,n) \). In Fig.~\ref{fig:ratio_plots_hydro_coeff}, we also illustrate the effect of shear viscosity on \( \AC_{2,1}(2,3) \). We found that for \( \AC_{2,1}(2,3) \), the observed variation is around 40-60\% across different centrality ranges; however, we note that the 0-10\% and 10-20\% centrality classes are subject to substantial statistical uncertainties. Additionally, the medium's bulk viscosity similarly impacts these observables. 

The influence of transport coefficients is notably more prominent for \(\AC_{2,1}(2,4)\) compared to \(\AC_{2,1}(2,3)\). Compared to traditional flow observables, asymmetrical cumulants display a heightened sensitivity to transport coefficients.

To further investigate these dynamics, we evaluated the sensitivities of the normalized asymmetric cumulants. From Fig.~\ref{fig:ratio_plots_hydro_coeff}, it becomes evident that \( \NAC_{2,1}(2,3) \) exhibits little sensitivity to shear viscosity. In contrast, \(\NAC_{2,1}(2,4)\) displays a significantly greater sensitivity to both shear and bulk viscosities compared to \( v_2\{4\} \).

\subsection{Sensitivity to Hadronic Interactions}
\begin{figure}[t!]
\includegraphics[width=0.45\textwidth]{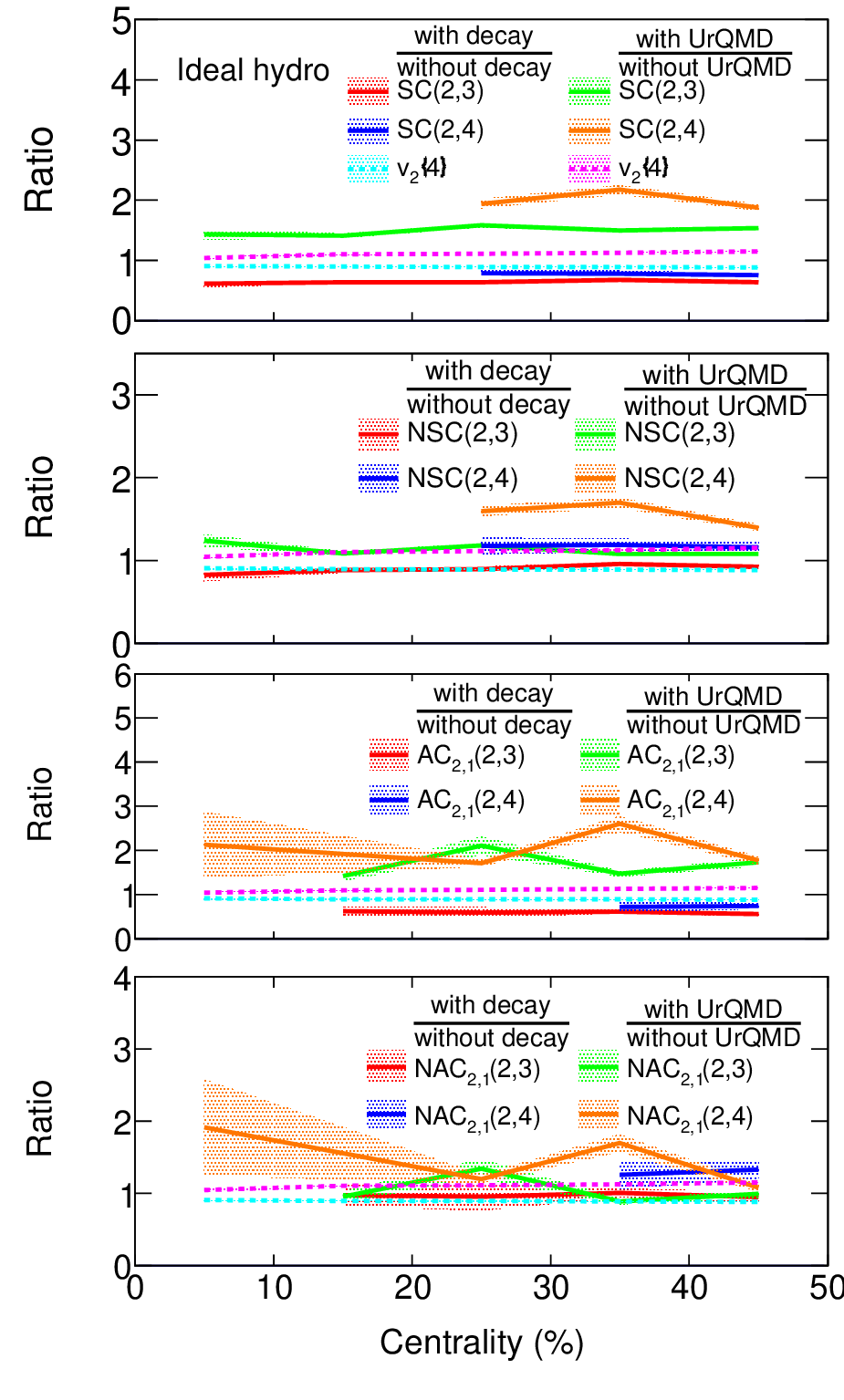}
\caption{\label{fig:ratio_plots_stages_ideal_hydro} Ratios of $v_2\{4\}$, $\SC(2,n)$, $\NSC(2,n)$, $\AC_{2,1}(2,n)$, and $\NAC_{2,1}(2,n)$, computed at the three different stages of the evolution, vs centrality in Au+Au collisions at $\sqrt{s_{NN}}$ = 200 GeV. The figure compares the effects of resonance decay and hadronic interactions on $v_2\{4\}$, $\SC(2,n)$, $\NSC(2,n)$, $\AC_{2,1}(2,n)$, and $\NAC_{2,1}(2,n)$. The plotted results show ratios of $v_2\{4\}$ using cyan (resonance decay) and magenta (hadronic interactions) bands and $\SC(2,n)$, $\NSC(2,n)$, $\AC_{2,1}(2,n)$, and $\NAC_{2,1}(2,n)$ using red (n=3, resonance decay), green (n=3, hadronic interactions), blue (n=4, resonance decay) and orange (n=4, hadronic interactions) bands. In the legends, the left column show the ratios for checking the effect of resonance decay and the right column show the ratios for checking the effect of hadronic interactions.}
\end{figure}
\begin{figure}[t!]
\includegraphics[width=0.45\textwidth]{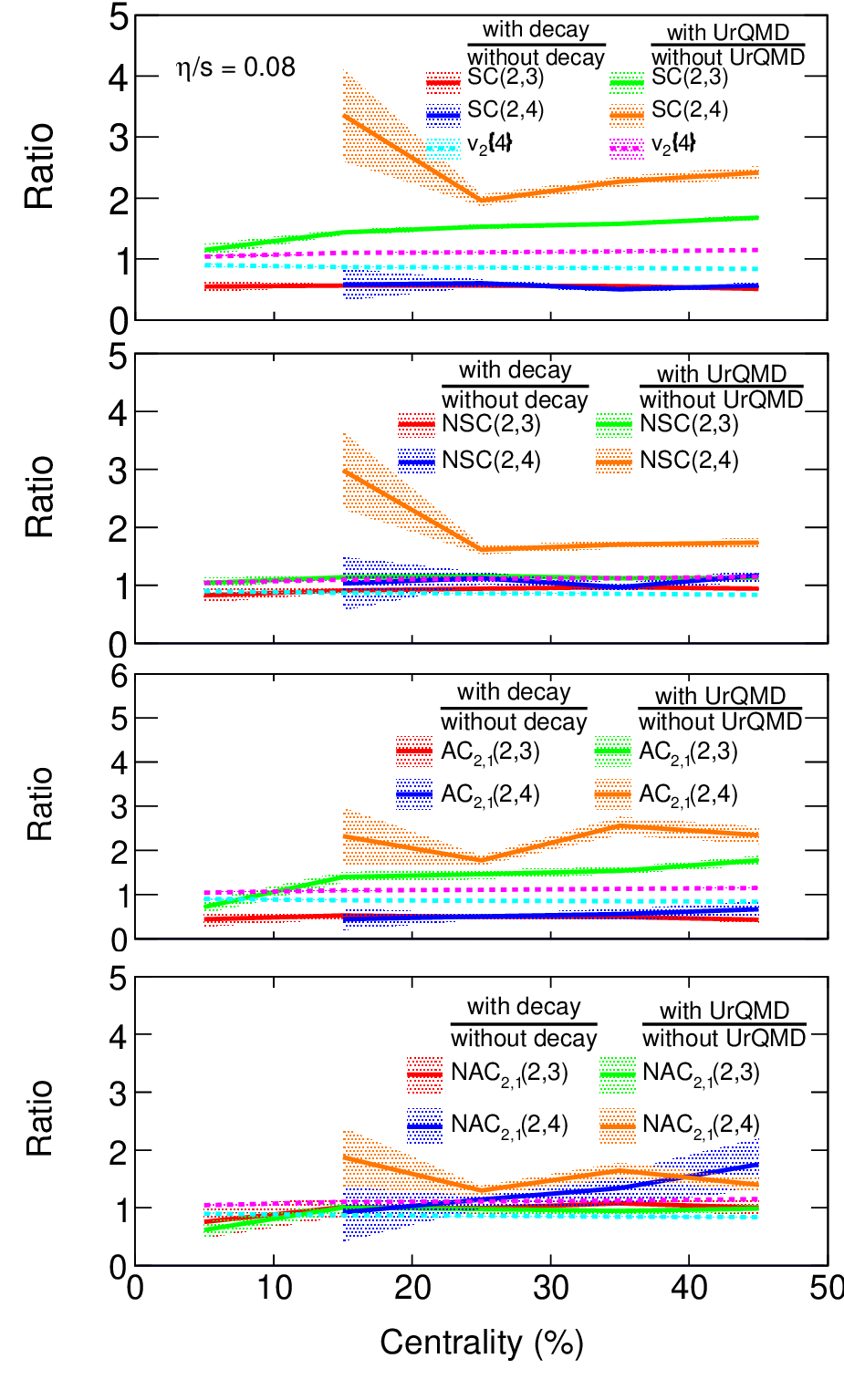}
\caption{\label{fig:ratio_plots_stages_etabys_0.08_zetabys_off} Ratios of $v_2\{4\}$, $\SC(2,n)$, $\NSC(2,n)$, $\AC_{2,1}(2,n)$, and $\NAC_{2,1}(2,n)$, computed at the three different stages of the evolution, vs centrality in Au+Au collisions at $\sqrt{s_{NN}}$ = 200 GeV. The figure compares the effects of resonance decay and hadronic interactions on $v_2\{4\}$, $\SC(2,n)$, $\NSC(2,n)$, $\AC_{2,1}(2,n)$, and $\NAC_{2,1}(2,n)$. The plotted results show ratios of $v_2\{4\}$ using cyan (resonance decay) and magenta (hadronic interactions) bands and $\SC(2,n)$, $\NSC(2,n)$, $\AC_{2,1}(2,n)$, and $\NAC_{2,1}(2,n)$ using red (n=3, resonance decay), green (n=3, hadronic interactions), blue (n=4, resonance decay) and orange (n=4, hadronic interactions) bands. In the legends, the left column show the ratios for checking the effect of resonance decay and the right column show the ratios for checking the effect of hadronic interactions.}
\end{figure}
\begin{figure}[t!]
\includegraphics[width=0.45\textwidth]{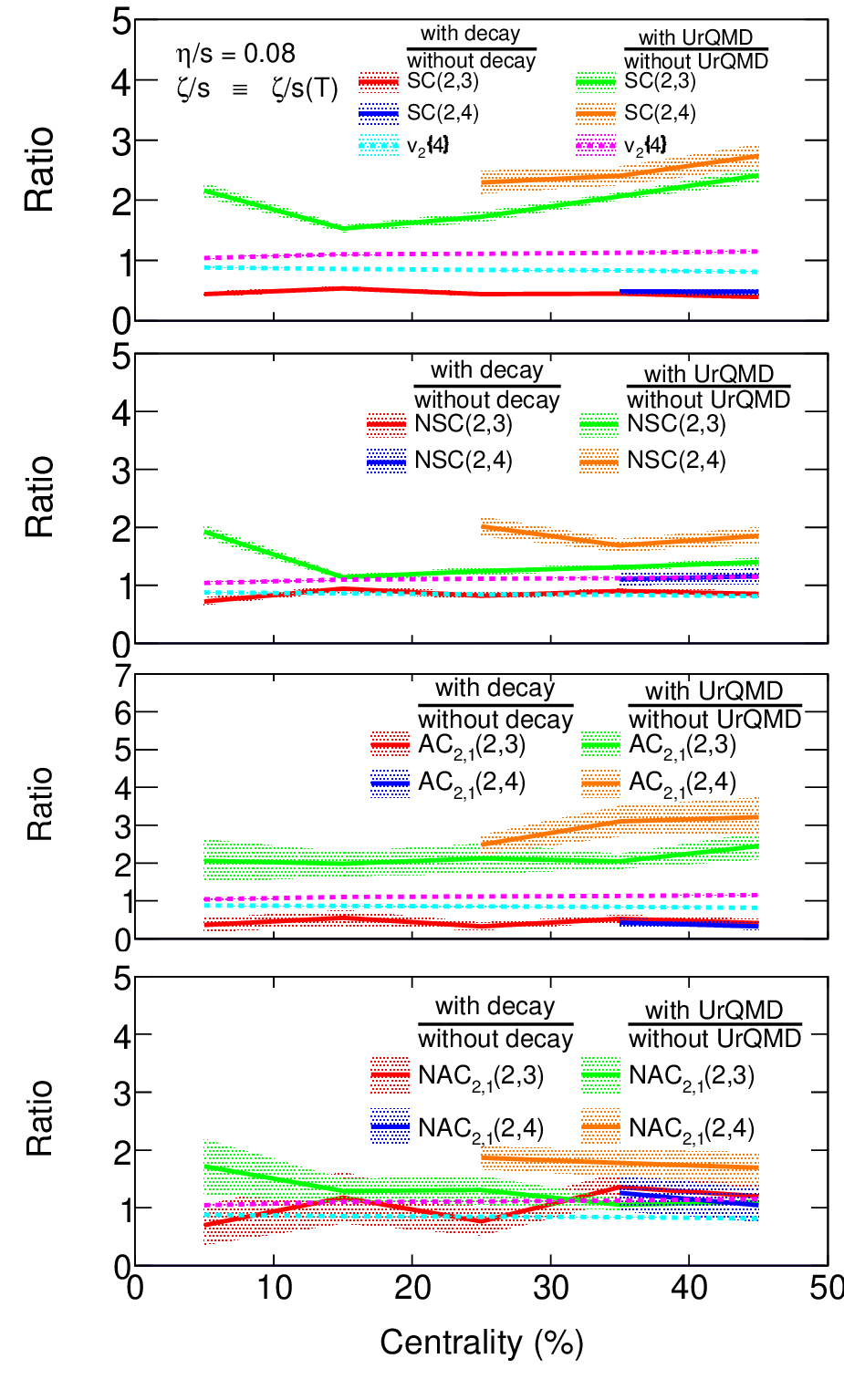}
\caption{\label{fig:ratio_plots_stages_etabys_0.08_zetabys_on} Ratios of $v_2\{4\}$, $\SC(2,n)$, $\NSC(2,n)$, $\AC_{2,1}(2,n)$, and $\NAC_{2,1}(2,n)$, computed at the three different stages of the evolution, vs centrality in Au+Au collisions at $\sqrt{s_{NN}}$ = 200 GeV. The figure compares the effects of resonance decay and hadronic interactions on $v_2\{4\}$, $\SC(2,n)$, $\NSC(2,n)$, $\AC_{2,1}(2,n)$, and $\NAC_{2,1}(2,n)$. The plotted results show ratios of $v_2\{4\}$ using cyan (resonance decay) and magenta (hadronic interactions) bands and $\SC(2,n)$, $\NSC(2,n)$, $\AC_{2,1}(2,n)$, and $\NAC_{2,1}(2,n)$ using red (n=3, resonance decay), green (n=3, hadronic interactions), blue (n=4, resonance decay) and orange (n=4, hadronic interactions) bands. In the legends, the left column show the ratios for checking the effect of resonance decay and the right column show the ratios for checking the effect of hadronic interactions.}
\end{figure}
\subsubsection{Four-Particle Symmetric Cumulants}
The hadronic afterburner is pivotal for interpreting data collected at the Relativistic Heavy Ion Collider (RHIC) \cite{Schenke:2019ruo}. To evaluate the sensitivity to late-stage hadronic interactions and resonance decay, we computed various observables at the conclusion of the three stages of evolution, as detailed in Table~\ref{table:stages}. Our analysis reveals that both $\SC(2,3)$ and $\SC(2,4)$ experience suppression due to resonance decay, while hadronic interactions lead to enhancement. Notably, the magnitudes of these effects exhibit significant differences between the two observables. Although the mean values show some consistency across different centralities, we exclude results for certain centralities for clarity, due to significant uncertainties.

Figs.~[\ref{fig:ratio_plots_stages_ideal_hydro}], [\ref{fig:ratio_plots_stages_etabys_0.08_zetabys_off}], and [\ref{fig:ratio_plots_stages_etabys_0.08_zetabys_on}] illustrate the impact of resonance decay and hadronic interactions on $v_2\{4\}$ and $\SC(2,3)$. For $v_2\{4\}$, we observe a variation of approximately 15\% due to resonance decay, and a 5\% change attributable to the interplay between hadronic interactions and resonance decay in mid-central collisions. In contrast, $\SC(2,3)$ displays a change of roughly 40\% driven by resonance decay across the full centrality range and a change of around 50\% from the combined effects of hadronic interactions and resonance decay. Furthermore, $\SC(2,4)$ is found to be more significantly affected by hadronic interactions compared to $\SC(2,3)$. Consequently, these observables reveal enhanced sensitivity to resonance decay and hadronic interactions.

Next, we explore the sensitivities of the normalized symmetric cumulants, which are expected to reduce the incidental sensitivity of \(\SC(m,n)\) to resonance decay and late-stage hadronic interactions. From Figs.~[\ref{fig:ratio_plots_stages_ideal_hydro}], [\ref{fig:ratio_plots_stages_etabys_0.08_zetabys_off}], and [\ref{fig:ratio_plots_stages_etabys_0.08_zetabys_on}], it becomes apparent that \(\NSC(2, 3)\) shows minimal sensitivity to resonance decay or late-stage hadronic interactions (UrQMD). Conversely, \(\NSC(2, 4)\) demonstrates a heightened sensitivity to late-stage hadronic interactions compared to $v_2\{4\}$.

\subsubsection{Six-Particle Asymmetric Cumulants}
We now present the results for $\AC_{2,1}(m,n)$. Figs.~[\ref{fig:ratio_plots_stages_ideal_hydro}], [\ref{fig:ratio_plots_stages_etabys_0.08_zetabys_off}], and [\ref{fig:ratio_plots_stages_etabys_0.08_zetabys_on}] display the influences of resonance decay and hadronic interactions on $\AC_{2,1}(m,n)$. For $\AC_{2,1}(2,3)$, we observe a noteworthy change due to these effects in mid-central collisions. The effects of late-stage hadronic interactions are observed to be more pronounced on $\AC_{2,1}(2,4)$ than on $\AC_{2,1}(2,3)$.

We also investigate the sensitivities of the normalized asymmetric cumulants. From Figs.~[\ref{fig:ratio_plots_stages_ideal_hydro}], [\ref{fig:ratio_plots_stages_etabys_0.08_zetabys_off}], and [\ref{fig:ratio_plots_stages_etabys_0.08_zetabys_on}], it is evident that \(\NAC_{2,1}(2, 3)\) exhibits minimal sensitivity to either resonance decay or late-stage hadronic interactions (UrQMD). In contrast, \(\NAC_{2,1}(2, 4)\) illustrates substantially greater sensitivity to these factors compared to $v_2\{4\}$.

As a result, the normalized (a)symmetric cumulant measurements possess the capability to distinguish between diverse models of quark-gluon plasma (QGP) evolution, particularly within hydrodynamic and transport frameworks. \(\NSC(2,3)\) and \(\NAC_{2,1}(2,3)\), due to their insensitivity to hydrodynamic model parameters, resonance decay, and late-stage hadronic interactions, are especially valuable for constraining the initial conditions of the system's evolution.

\subsection{Centrality dependence of \(\SC(m,n)\) and \(\AC_{2,1}(m,n)\)}
\subsubsection{Four-particle symmetric cumulants}
\begin{figure}[t!]
\includegraphics[width=0.45\textwidth]{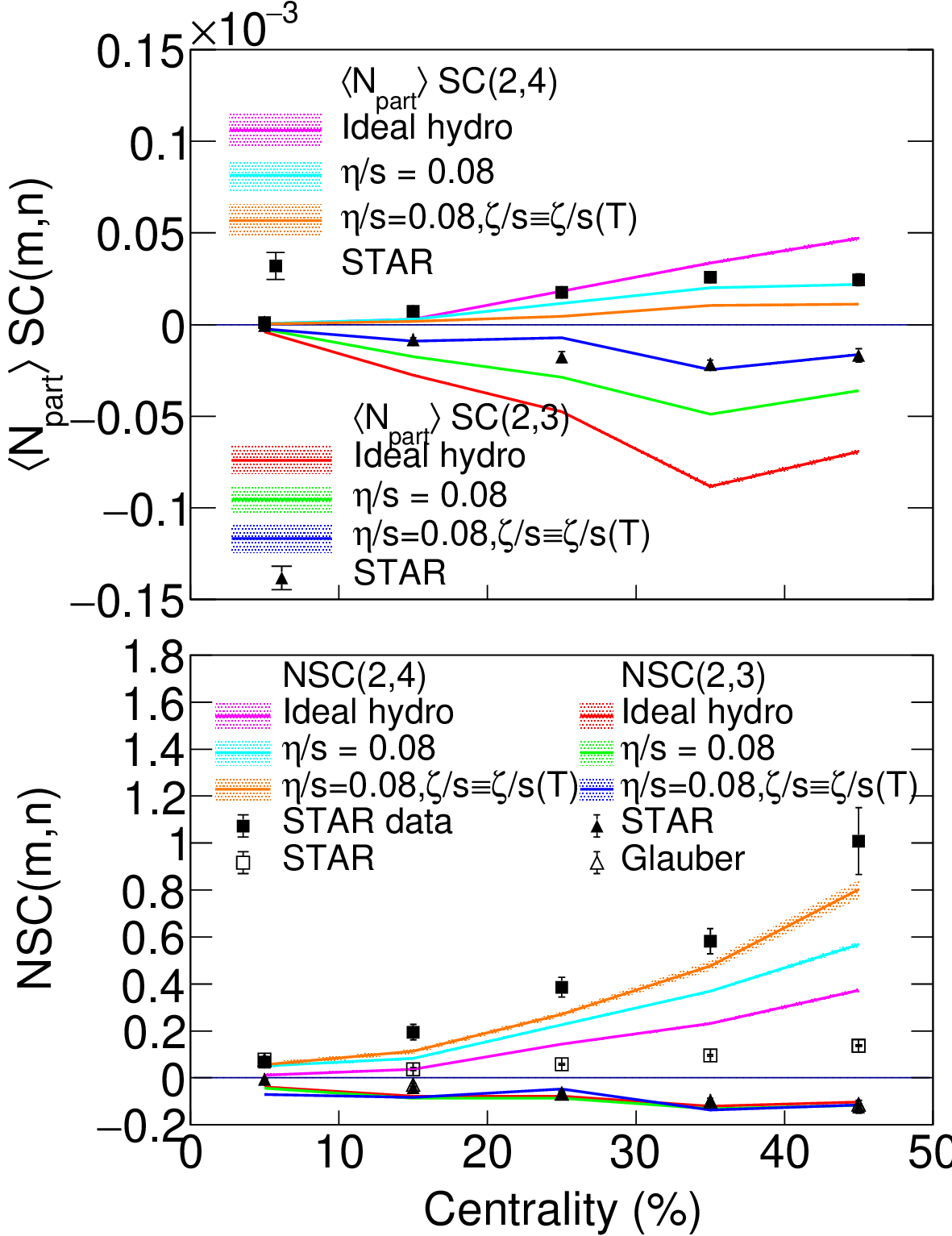}
\caption{\label{fig:N_SCvsCen} (Normalized) Symmetric cumulant vs centrality from four-particle correlations compared with STAR data~\cite{STAR:2018fpo} in Au+Au collisions at $\sqrt{s_{NN}}$ = 200 GeV. $\SC(2, 3)$ is consistently negative while as $\SC(2, 4)$ is consistently positive. Note that we have shown the Glauber results for $\NSC(2,4)$ to highlight that in non-central collisions $v_4$ is mainly driven by $\varepsilon_{2}^{2}$ and not $\varepsilon_{4}$.}
\end{figure}

Next, we compare the results from our simulations with the experimental measurements from RHIC. Fig. [\ref{fig:N_SCvsCen}] shows comparisons between the theory calculations for the symmetric cumulants $\SC(2, 3)$ and $\SC(2, 4)$ multiplied by $\langle N_{part}\rangle$ and the same measured by the STAR Collaboration~\cite{STAR:2018fpo}. The factor of $\langle N_{part}\rangle$ is multiplied to scale out the trivial dilution of correlation with the increase in the number of multiplets. The magnitudes of $\langle N_{part}\rangle\SC(2, 3)$ and $\langle N_{part}\rangle\SC(2, 4)$ increase from central to mid-peripheral events and then again decreases for very peripheral events. Here, we do not aim to reproduce the experimental data because our aim is to check the sensitivity of these observables to the transport properties of the system created in these collisions. The comparison in Fig.~[\ref{fig:N_SCvsCen}] reflects that the correlation in initial spatial eccentricities and the following hydrodynamic evolution can capture the correlation among flow harmonics of different orders. Negative values of $\SC(2, 3)$ throughout the centrality range reveal the anti-correlation between $v_2$ and $v_3$. While as $\SC(2, 4)$ is positive for all centralities, indicating positive correlations between $v_2$ and $v_4$.
Symmetric cumulants do not have contributions from non-flow effects, where non-flow refers to azimuthal correlations not related to the reaction plane orientation, like those from resonances, jets, quantum statistics, etc. This is verified by computing these observables for the HIJING model, which includes only non-flow physics, for which these are consistently zero~\cite{PhysRevC.105.024912}. The model reproduces the qualitative variation of $\SC(m,n)$ with the centrality. However, a single set of transport coefficients cannot simultaneously explain both $\SC(2, 3)$ and $\SC(2, 4)$.
\par
The normalized symmetric cumulants in Au+Au collisions at $\sqrt{s_{NN}}$ = 200 GeV, computed using Eq.~[\ref{scmn_norm1}] in momentum space and the corresponding equation in coordinate space for the initial state eccentricity, $\varepsilon_{n}$ (Eq.~[\ref{eq_e_n}]), evaluated using the Monte Carlo Glauber model, are presented in Fig.~[\ref{fig:N_SCvsCen}].
\begin{eqnarray}
    \label{eq_e_n}
    \varepsilon_n e^{in\psi_n} &=& -\frac{\int r^n e^{in\phi} \rho\left(r,\phi\right) rdrd\phi}{\int r^n \rho\left(r,\phi\right)rdrd\phi}, \hspace{1mm}n\geq2
\end{eqnarray}
We observe that $\NSC(2,3)$ shows little sensitivity to shear and bulk viscosities. However, $\NSC(2,4)$ is more sensitive to both shear and bulk viscosities than $v_2\{4\}$. This is due to the fact that \( v_4 \) is the superposition of a linear term and a nonlinear term. This can be expressed as~\cite{Teaney:2012ke}:
\begin{equation}
v_{4}e^{-i4\psi_{4}} = w_{4}e^{-i4\phi_{4}} + w_{4(22)}e^{-i4\phi_{2}},
\end{equation}
where \(w_{4}\) represents the linear response while \(w_{4(22)}\) describes the nonlinear response. Notably, the linear term, \( w_{4} \), experiences greater damping compared to the nonlinear term, \(w_{4(22)}\), as noted in Reference~\cite{Teaney:2012ke}. Furthermore, \( w_{4} \) demonstrates a weaker correlation with \( v_2 \). Consequently, as viscosity increases, the correlation between \( v_4 \) and \( v_2 \) also increases~\footnote{We thank the anonymous referee for their insightful comments and suggestions, particularly the detailed explanation that helped us expand the context of this result.}.
\par
If only eccentricity drives $v_n$, then we can expect that the $\NSC(m,n)$ in the final state would be equal to the $\NSC(m,n)$ in the initial state. Fig.~[\ref{fig:N_SCvsCen}] shows that the initial anti-correlation between $\varepsilon_2$ and $\varepsilon_3$ is mainly responsible for the observed anti-correlation between $v_2$ and $v_3$. However, the correlation between $\varepsilon_2$ and $\varepsilon_4$ is smaller than the observed correlation between $v_2$ and $v_4$. The contribution to anisotropic flow $v_4$ not only comes from the linear response of the system to $\varepsilon_4$ but also has a contribution proportional to $\varepsilon_2^2$. As collisions become more peripheral, the difference between $\NSC(2,4)$ in the final state and the initial state increases, likely due to $\varepsilon_2$ playing a more significant role in $v_4$. This has also been observed in Pb+Pb collisions at $\sqrt{s_{NN}}$ = 2.76 TeV by the ATLAS~\cite{ATLAS:2015qwl} and ALICE~\cite{ALICE:2016kpq} experiments. The properties of the medium were suggested to affect the relative contribution of $\varepsilon_2$ in $v_4$ compared to $\varepsilon_4$~\cite{Teaney:2012ke}. Therefore, $\NSC(2,4)$ provides a probe into the medium properties.

\subsubsection{Six-particle asymmetric cumulants}
\begin{figure}[t!]
\includegraphics[width=0.45\textwidth]{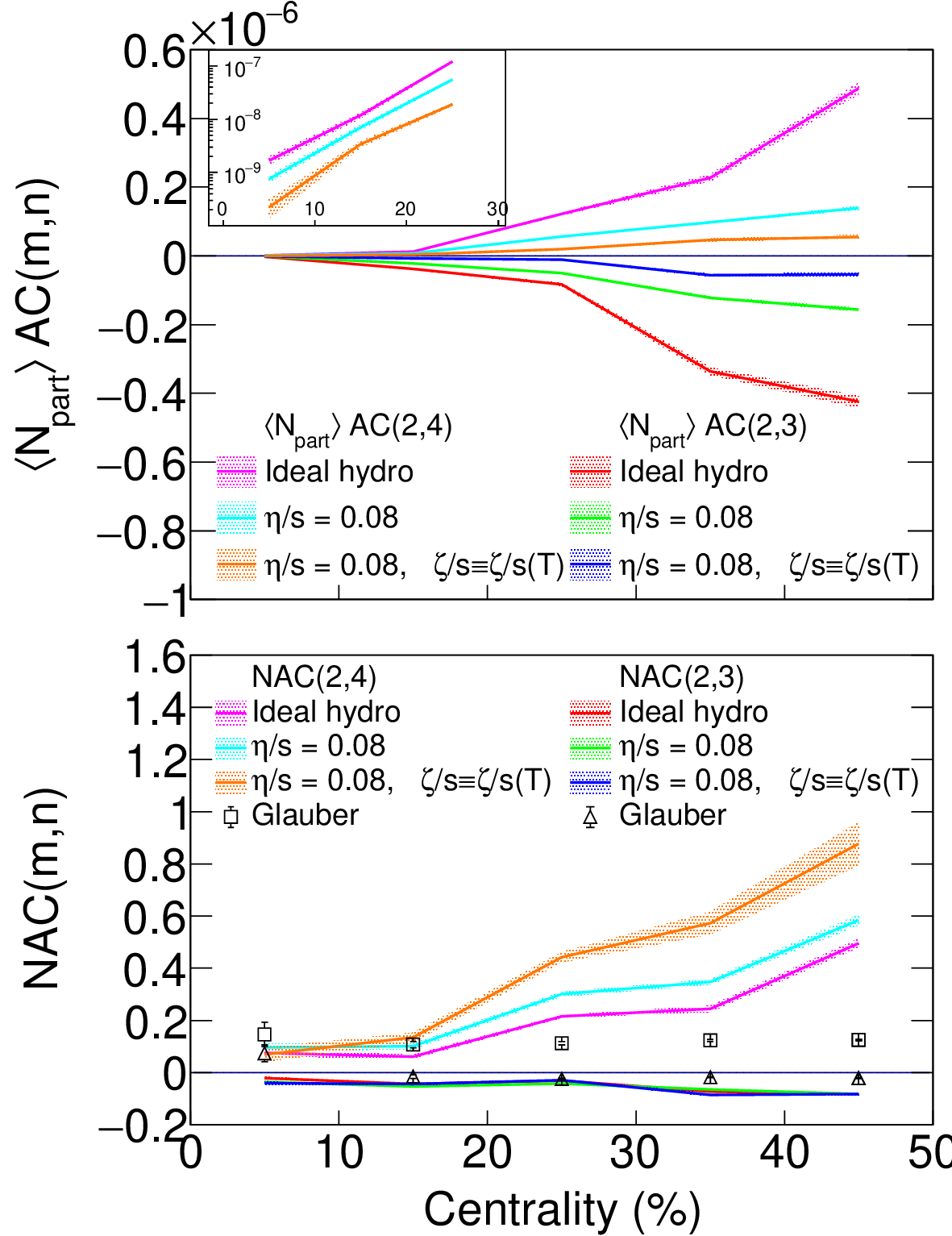}
\caption{\label{fig:N_ASC23ASC24vsCen} (Normalized) Asymmetric cumulant vs centrality from six-particle correlations in Au+Au collisions at $\sqrt{s_{NN}}$ = 200 GeV. $\AC_{2,1}(2, 3)$ is consistently negative while as $\AC_{2,1}(2, 4)$ is consistently positive. Note that we have shown the Glauber results for $\NAC_{2,1}(2,4)$ to highlight that in non-central collisions $v_4$ is mainly driven by $\varepsilon_{2}^{2}$ and not $\varepsilon_{4}$.}
\end{figure}

In Fig.~[\ref{fig:N_ASC23ASC24vsCen}], we show the centrality dependence of $\AC_{2,1}(m,n)$, multiplied by $\langle N_{part}\rangle$, for Au+Au at $\sqrt{s_{NN}}=200 \ GeV$. The magnitudes of $\langle N_{part}\rangle\AC_{2,1}(2,3)$ and $\langle N_{part}\rangle\AC_{2,1}(2,4)$ increase with centrality. As for symmetric cumulants, negative values of $\AC_{2,1}(2, 3)$ throughout the centrality range reveal the anti-correlation between $v_2$ and $v_3$. While as $\AC_{2,1}(2, 4)$ is positive for all centralities, indicating positive correlations between $v_2$ and $v_4$. We compare these observables for the three sets of ensembles mentioned in Table ~\ref{table:parsets} and observe that both $\AC_{2,1}(2,3)$ and $\AC_{2,1}(2,4)$ are suppressed by shear and bulk viscosities.
\par
In Fig.~[\ref{fig:N_ASC23ASC24vsCen}], we have compared the normalized asymmetric cumulant in coordinate and momentum space, as computed using Eq.~[\ref{NACDefinition}] and the corresponding equation for the initial state eccentricities (Eq.~[\ref{eq_e_n}]). Again, if only eccentricity drives $v_n$, we can expect the $\NAC_{2,1}(m,n)$ in the final state to be equal to the $\NAC_{2,1}(m,n)$ in the initial state. Fig. [\ref{fig:N_ASC23ASC24vsCen}] illustrates that the initial negative correlation between $\varepsilon_2$ and $\varepsilon_3$ primarily causes the observed negative correlation between $v_2$ and $v_3$. However, unlike symmetric cumulants, here, the initial state correlations are smaller in the peripheral collisions. Similar to the reasons mentioned for $\NSC(2,4)$, as collisions become more peripheral, the difference in $\NAC_{2,1}(2,4)$ between the initial and the final states increases. We compare these observables for the three sets of ensembles mentioned in Table ~\ref{table:parsets} and observe that, while $\NAC_{2,1}(2,3)$ remains unaffected, $\NAC_{2,1}(2,4)$ is notably enhanced by the influence of shear and bulk viscosities. As discussed earlier, the linear component in \(v_4\) is damped more significantly than the nonlinear term and shows a weaker correlation with \(v_2\). Consequently, as viscosity increases, the correlation between \(v_4\) and \(v_2\), and thus $\NAC_{2,1}(2,4)$, is also enhanced.
\par
Currently, STAR has not yet measured asymmetric cumulants. We conducted a comparison between our findings and the Pb+Pb 5.02 TeV results from ALICE 
(See Fig.~\ref{fig:SC_AC_STAR_vs_ALICE} in Appendix~\ref{app1}). The centrality dependence of the observables we discussed in this paper shows similarities between ALICE and STAR data. The \(\SC(m,n)\) values are about an order of magnitude larger at ALICE than at STAR. Similarly, the \(\AC_{2,1}(m,n)\) values are 3 to 5 times larger at ALICE compared to STAR. However, the similarities in the magnitude and centrality dependence of the normalized cumulants are noteworthy. This comparison indicates minimal energy dependence for the normalized cumulants.

\section{Conclusions}
We have presented results for multi-particle correlation functions in heavy-ion collisions at top RHIC energy using a hybrid framework based on the Monte Carlo Glauber model, MUSIC viscous hydrodynamics simulations, and the UrQMD hadronic cascade.
\par
First, we tuned the free parameters, such as shear and bulk viscosities, to describe particle multiplicities, mean transverse momentum, and anisotropic flow. Subsequently, we discussed the impact of shear and bulk viscosities, resonance decay, and hadronic interactions on traditional and novel observables.
\par
We have studied correlators that measure the correlations between flow harmonics of varying orders, specifically four and six-particle correlations, including both symmetric and asymmetric cumulants as functions of centrality at midrapidity in Au+Au collisions at $\sqrt{s_{NN}}=200$ GeV. 
These observables are highly sensitive to shear and bulk viscosities, resonance decay, and hadronic interactions. However, this magnified sensitivity can be attributed to the properties of anisotropic flow, particularly as the cumulants incorporate higher powers of flow. We then examined the sensitivities of the normalized cumulants. We found that \(\NSC(2, 3)\) and \(\NAC_{2,1}(2, 3)\) are insensitive to both the hydro model parameters (shear and bulk viscosity) and late-stage hadronic interactions. This strongly supports their effectiveness as reliable tools for constraining the initial state of the system’s evolution. Conversely, \(\NSC(2,4)\) and \(\NAC_{2,1}(2,4)\) show considerable sensitivity to these stages, allowing them to effectively constrain hydrodynamic and transport models.
\par
We observed anti-correlation between event-by-event fluctuations of $v_2$ and $v_3$, while the event-by-event fluctuations of $v_2$ and $v_4$ are found to be positively correlated. The observed anti-correlation between $v_2$ and $v_3$ appears to be described by the initial-stage anti-correlation between $\varepsilon_{2}$ and $\varepsilon_{3}$, which supports the idea of linearity between $\varepsilon_{n}$ and $v_n$~\cite{Alver:2010gr}. The hydrodynamic response suggests that the final state fluctuation originates from the initial state. However, it is important to consider the nonlinear hydrodynamic response of the medium to explain the measured correlation between $v_2$ and $v_4$, as the initial-stage linear correlation alone is insufficient. 
\par
The values of \(\SC(m,n)\) and \(\AC_{2,1}(m,n)\) at ALICE are larger—by factors ranging from 3 to 8—compared to those at STAR. However, it is important to note the similarities in the magnitude and centrality dependence of the normalized cumulants. This comparison indicates that there is minimal energy dependence for the normalized cumulants.

\section{Acknowledgements}
We would like to acknowledge Tribhuban Parida for fruitful discussions and providing us the computational setup. We would like to express our sincere gratitude to the anonymous referee for their invaluable insights and constructive feedback, especially for the thorough and thoughtful explanation they provided regarding one of our key results.

\bibliography{main}

\section{Appendix}
\label{app1}
\begin{figure*}
\includegraphics[width=0.45\textwidth]{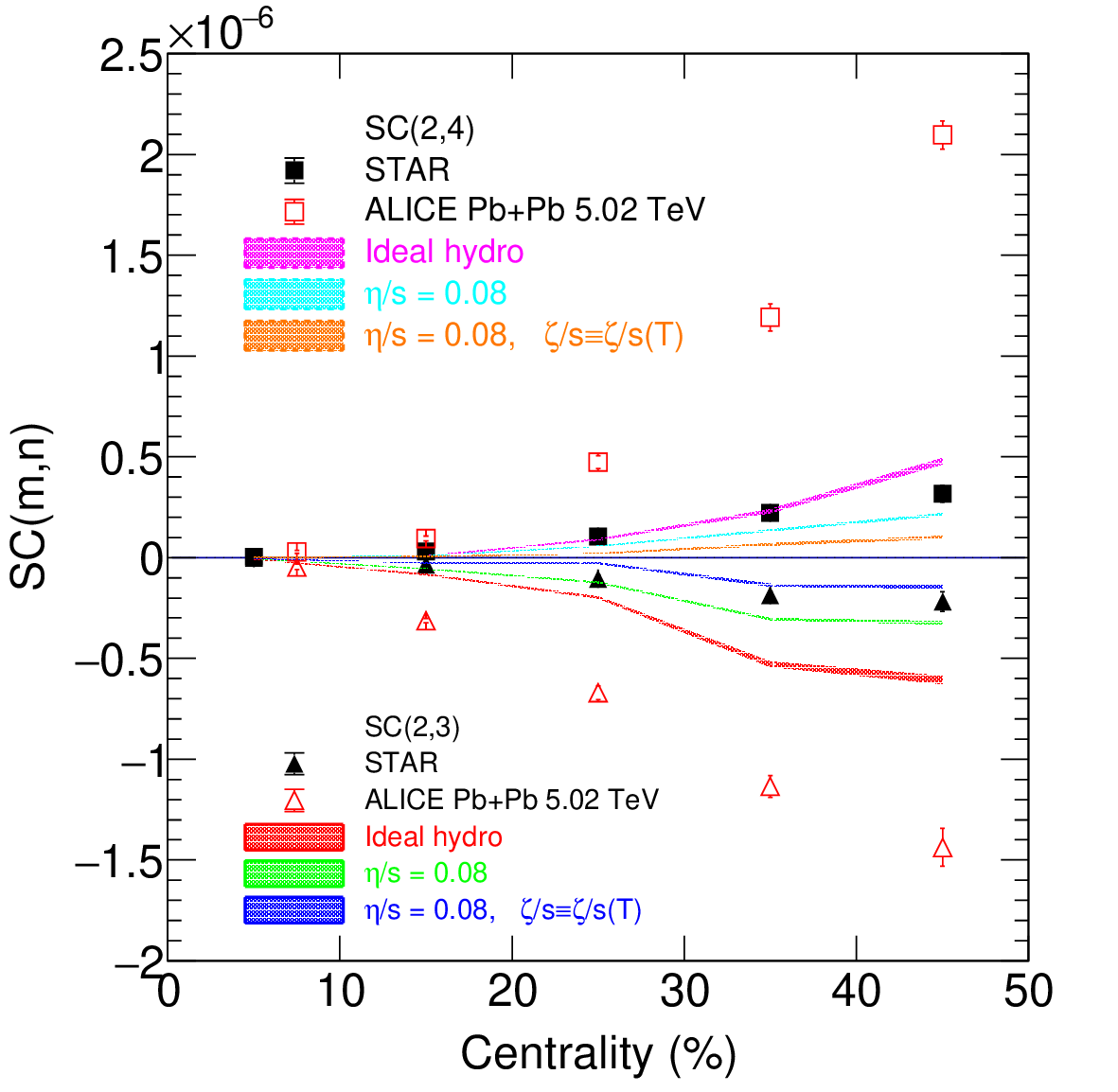}
\includegraphics[width=0.45\textwidth]{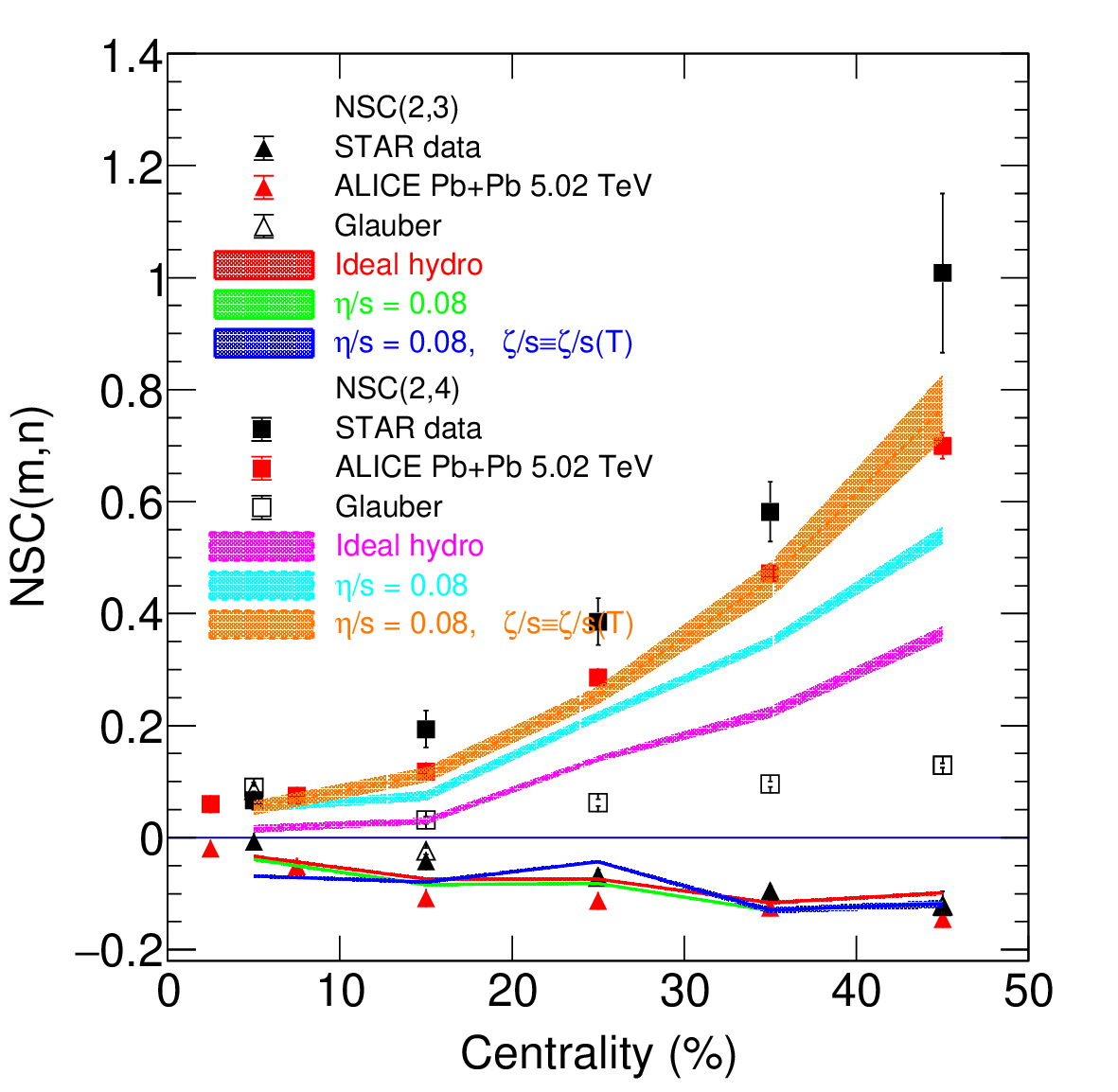}
\includegraphics[width=0.45\textwidth]{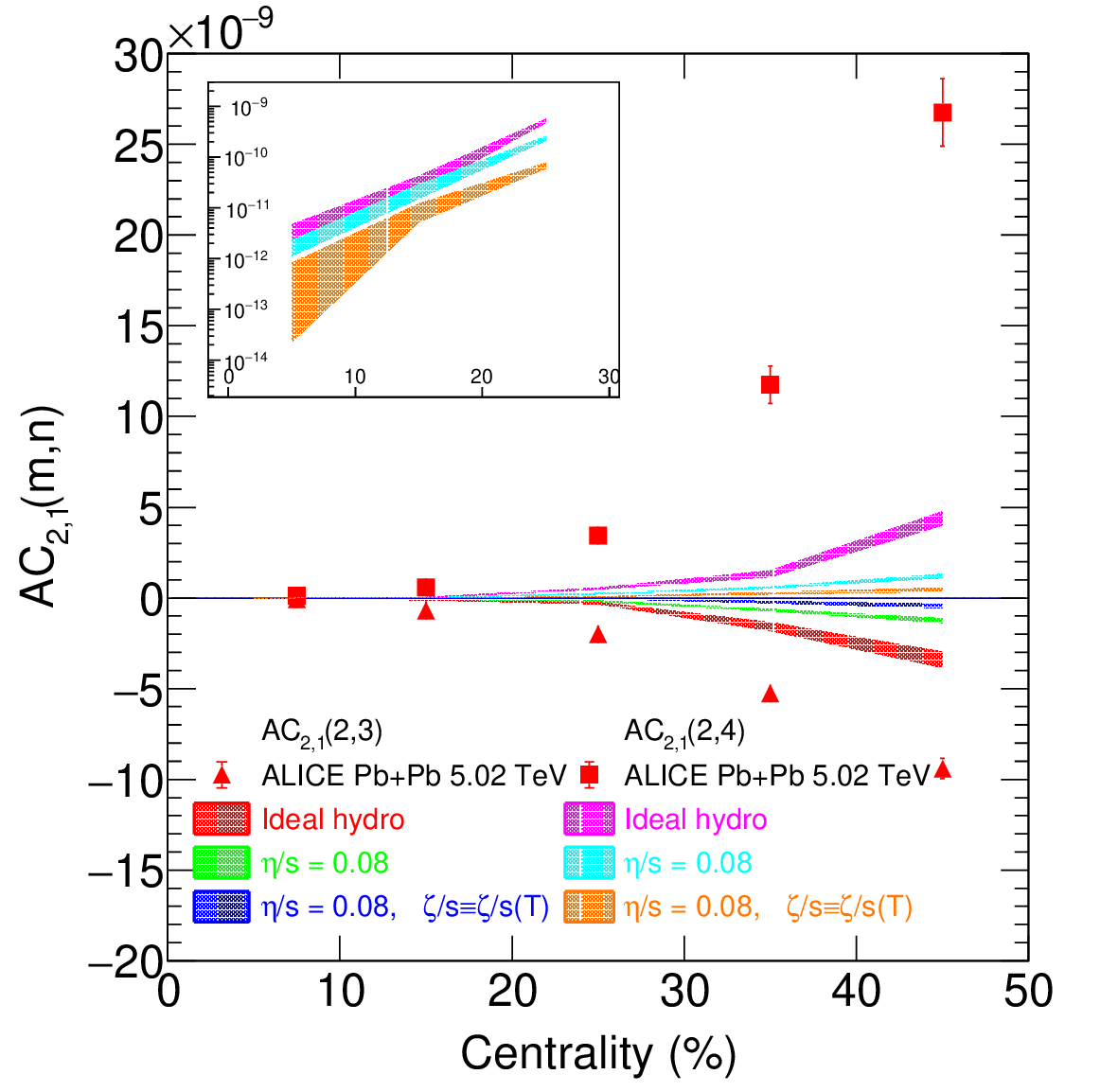}
\includegraphics[width=0.45\textwidth]{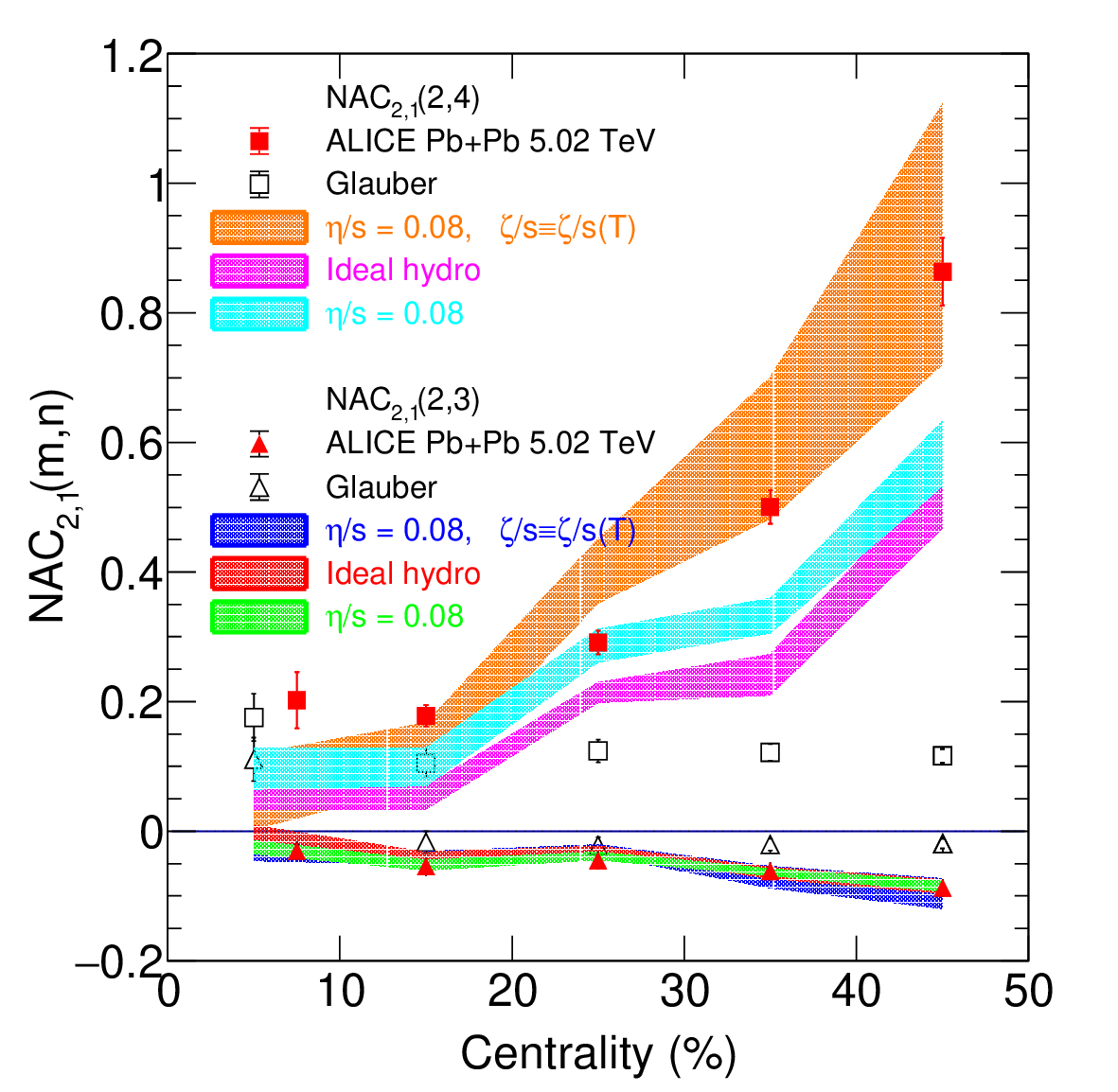}
\caption{\label{fig:SC_AC_STAR_vs_ALICE}
Comparison of $\SC(m, n)$ (top-left), $\NSC(m, n)$ (top-right), $\AC_{2,1}(m,n)$ (bottom-left), and $\NAC_{2,1}(m,n)$ (bottom-right) as functions of centrality in Au+Au collisions at $\sqrt{s_{NN}} = 200$~GeV. 
The symmetric cumulants $\SC(m, n)$ and $\NSC(m, n)$ are obtained from four-particle correlations, while the asymmetric cumulants $\AC_{2,1}(m,n)$ and $\NAC_{2,1}(m,n)$ are computed from six-particle correlations. 
Results are compared with STAR measurements~\cite{STAR:2018fpo} (for symmetric cumulants) and ALICE measurements in Pb+Pb collisions at 5.02~TeV~\cite{ALICE:2023lwx,ALICE:2021adw}.
}
\end{figure*}

Fig.~\ref{fig:SC_AC_STAR_vs_ALICE} provides a visual comparison between our results, STAR~\cite{STAR:2018fpo} and previously published Pb+Pb 5.02~TeV data from the ALICE~\cite{ALICE:2023lwx,ALICE:2021adw} collaboration. While STAR has not yet reported asymmetric cumulants, this comparison allows us to examine the qualitative and quantitative behavior of these observables across different energies and collision systems. The $\SC(m,n)$ values are observed to be approximately an order of magnitude larger in ALICE data compared to STAR. Similarly, the $\AC_{2,1}(m,n)$ values are typically 3 to 5 times greater at ALICE. Despite these differences in absolute magnitude, the normalized cumulants demonstrate similar centrality trends and relative scaling between the two experiments. These figures highlight that normalized cumulants show minimal energy dependence, which supports the interpretation presented in the main text.

\end{document}